\documentclass[]{aa}  

\usepackage{multicol}
\usepackage{graphicx}
\usepackage{multirow}
\usepackage{amsmath}
\usepackage{txfonts}
\usepackage{natbib}
\bibpunct{(}{)}{;}{a}{}{,} 
\def\ltsim{\raise 2pt \hbox {$<$} \kern-1.1em \lower 4pt \hbox {$\sim$}}
\def\gtsim{\raise 2pt \hbox {$>$} \kern-1.1em \lower 4pt \hbox {$\sim$}}

\begin{document} 

\title{
A LOFAR view into the stormy environment of the galaxy cluster 2A0335+096}

\author{A. Ignesti\inst{1,2}\thanks{Contact: {\tt alessandro.ignesti@inaf.it}}, 
G. Brunetti\inst{3},
T. Shimwell\inst{4,5},
M. Gitti\inst{2,3}, 
L. Birzan\inst{6},
A. Botteon\inst{5}, 
M. Br\"uggen\inst{6}, 
F. de Gasperin\inst{6,3},
G. Di Gennaro\inst{6}, 
A. C. Edge\inst{7},
C. J. Riseley\inst{2,3,8},
H. J. A. R\"ottgering\inst{5}, 
R. J. van Weeren\inst{5}
}
\institute{
INAF-Padova Astronomical Observatory, Vicolo dell’Osservatorio 5, I-35122 Padova, Italy
\and
 Dipartimento di Fisica e Astronomia, Universit\`a di Bologna, via Gobetti 93/2, 40129 Bologna, Italy
\and 
INAF, Istituto di Radioastronomia di Bologna, via Gobetti 101, 40129 Bologna, Italy
\and
ASTRON, the Netherlands Institute for Radio Astronomy, Postbus 2, 7990 AA Dwingeloo, The Netherlands
\and
Leiden Observatory, Leiden University, PO Box 9513, 2300 RA Leiden, The Netherlands
\and
Hamburger Sternwarte, Universit\"at Hamburg, Gojenbergsweg 112, D-21029, Hamburg, Germany 
\and
Centre for Extragalactic Astronomy, Durham University, DH1 3LE, UK 
\and
CSIRO Space \& Astronomy, PO Box 1130, Bentley, WA 6102, Australia 
}

\authorrunning{Ignesti et al.}
\titlerunning{A LOFAR view into the stormy environment of the galaxy cluster 2A0335+096}

\date{Accepted }

\abstract 
{
 Radio observations represent a powerful probe of the physics occurring in the intracluster medium (ICM) because they trace the relativistic cosmic rays in the cluster magnetic fields, or within galaxies themselves. By probing the low-energy cosmic rays, low-frequency radio observations are especially interesting because they unveil emission powered by low-efficiency particle acceleration processes, which are believed to play a crucial role in the origin of diffuse radio emission. }
{
We investigate the origin of the radio mini-halo at the center of the galaxy cluster 2A0335+096 and its connection to the central galaxy and the sloshing cool core. We also study the properties of the head-tail galaxy GB6 B0335+096 hosted in the cluster to explore the lifecycle of the relativistic electrons in its radio tails.
}
{
We use new LOw Frequency ARray (LOFAR) observations from the LOFAR Two-meter Sky Survey at 144 MHz to map the low-frequency emission with a high level of detail. The new data were combined with archival Giant Metrewave Radio Telescope and $Chandra$ observations to carry out a multi-wavelength study. }
{
We have made the first measurement of the spectral index  of the mini-halo ($\alpha=-1.2\pm0.1$ between 144 MHz and 1.4 GHz) and the lobes of the central source ($\alpha\simeq-1.5\pm0.1$ between 144 and 610 MHz). Based on the low-frequency radio emission morphology with respect to the thermal ICM, we propose that the origin of the diffuse radio emission is linked to the sloshing of the cool core. The new data revealed the presence of a Mpc-long radio tail associated with GB6 B0335+096. The observed projected length is a factor 3 longer than the expected cooling length, with evidence of flattening in the spectral index trend along the tail. Therefore, we suggest that the electrons toward the end of the tail are kept alive by the ICM gentle re-acceleration.
 } {}


\maketitle
\section{Introduction}
Low-frequency ($\sim100$ MHz) observations probe low-energy particles that, due to their lower emissivity, have typical life-times of a factor $\sim3$ longer than the high-energy particles traced by high-frequency ($>$1000 MHz) radio emission \citep[e.g.,][]{Pacholczyk_1970}. This difference in time-scales is crucial when studying the re-acceleration processes of cosmic rays electrons (CRe). This is because a re-acceleration process, and re-energization in general, is effective only if the time-scale of the energy gain is smaller (or, at least, equal) than the time-scale of the energy losses. It follows that low-energy CRe, due to their longer lifetimes in a given magnetic field \citep[e.g., 1-10 $\mu$G for galaxy clusters, see][]{Carilli_2002}, are more prone to be affected by re-acceleration than the high-energy ones. Therefore, only low-frequency observations have the potential to explore these phenomena, hence opening a window on a class of low-efficiency processes related to the intracluster medium (ICM) microphysics, in form of turbulence or gentle compression of the radio plasma by the thermal gas \citep[for a review][and references therein]{Brunetti-Jones_2014}.\\

In this work we present an analysis of the low-frequency observation of the galaxy cluster 2A0335+096 carried out with the LOw Frequency ARray \citep[LOFAR,][]{vanHaarlem_2013}. 2A0335+096 is a low redshift ($z=0.035$),  low-mass cluster \citep[M$_{500}=2.27\times10^{14}$ M$_{\odot}$;][]{Planck_2014}\footnote{ M$_{500}$ and R$_{500}$ indicate the total mass and
radius at a mean over-density of 500 with respect to the cosmological critical density at redshift of the cluster, respectively. }. We selected this cluster on the basis of its interesting features. To begin with, 2A0035 hosts a radio mini-halo (MH) at its center \citep[][]{Sarazin_1995}. MHs are diffuse radio sources observed at the center of relaxed clusters and characterized by steep spectral indices ($-1.5<\alpha<-1.1$)\footnote{In this work we define the synchrotron spectrum as $S\propto \nu^{\alpha}$, where $S$ is the flux density at the frequency $\nu$ and $\alpha$ is the spectral index.} and typical radii of 100-150 kpc  \citep[e.g.,][and the references therein]{Giacintucci_2017, VanWeeren_2019}. Usually the radio emission is mostly confined within the cool-core region of the clusters, thus outlining a connection between the non-thermal ICM components and the thermal plasma \citep[e.g.,][]{Rizza_2000,Gitti_2002,Mazzotta-Giacintucci_2008,Giacintucci_2014b}. This intrinsic connection is supported by the correlations observed between integrated radio and X-ray luminosities \citep[][]{Bravi_2016,Gitti_2015,Gitti_2018,Giacintucci_2019,Richard_2020}. Recently, the discovery of very steep-spectrum ($\alpha<-2$), low-frequency emission on large scales ($\sim500$ kpc) around a number of MHs \citep[][]{Savini_2018,Savini_2019,Biava_2021} suggested the possibility that low-energy CRe could actually diffuse over such large scales also in relaxed clusters. The origin of CRe in the MH volume is still debated, with two proposed scenarios. One is the leptonic scenario, where the CRe, possibly injected by the active galactic nucleus (AGN) of the central radio galaxy, are re-accelerated by ICM turbulence \citep[e.g.,][]{Gitti_2002,ZuHone_2013}. The other scenario is the hadronic case, where CRe are produced by collisions between the relativistic and the thermal protons of the ICM, which can travel to larger distances due to their lower emissivity \citep[][]{Pfrommer-Ensslin_2004}.\\

The MH in 2A0335+096 was first imaged at 1.4 GHz and 5.5 GHz by \citet[][]{Sarazin_1995}. It surrounds the central radio galaxy and extends for only $\sim100''$ ($\sim$ 70 kpc), thus making it one of the smallest MH known \citep[][]{Giacintucci_2017,Giacintucci_2019}. The detection of extended radio emission at 144 MHz surrounding the central galaxy has already been presented in \citet[][]{Kokotanekov_2017}. The central radio galaxy presents two components: there is an inner source \citep[][]{Donahue_2007,Giacintucci_2019} and a pair of large-scale radio lobes filling the X-ray cavities \citep[][]{Birzan_2020}. There is also a third patch of extended emission, which is interpreted as a fossil lobe from an older AGN outburst, detected at $\sim 25 ''$($\sim$18 kpc) from the cluster center \citep[][]{Giacintucci_2019}. \citet[][]{Sarazin_1995} discussed the possibility that part of the radio emission comes also from a companion galaxy close to the cluster center. \\

From the thermal point of view, 2A0335+096 has a central temperature of 3.6 keV and a central entropy of 7.1 keV cm$^{-2}$ \citep[][]{Cavagnolo_2009,Giacintucci_2017}. In the X-ray band it presents two cavities, which coincide with the position of the radio lobes. In addition, there is a cold front located at $\sim$40 kpc from the cluster center, which suggests that the cool core is sloshing in the gravitaional well \citep[][]{Mazzotta_2003, Ghizzardi_2006, Sanders_2009b}.\\

Evidences of the stormy nature of 2A0335+096 come also from optical and ALMA observations. \citet[][]{Vantyghem_2016} and \citet[][]{Olivares_2019} present detailed studies of cold and warm filaments at the cluster center. The gas appears to be distributed over several clumpy and dusty filaments trailing the X-ray cavities, which support the idea that the gas has
cooled from low-entropy ICM that could have been lifted out of the cluster core, becoming thermally unstable. Both the molecular and the warm gas present disturbed  kinematics, mainly due to the projection of several inflowing or outflowing filaments, or due to the presence of turbulence induced by bubbles, jets, or sloshing.\\

The stormy environment in which we observe the MH, with the cavities inflated by the AGN and the sloshing, would entail a perturbed, turbulent environment, at least in the central region of the ICM and possibly beyond. Therefore, motivated by the recent findings of low-frequency diffuse emission extending beyond cool cores \citep[][]{Savini_2018,Savini_2019,Biava_2021}, we obtained 8.3 hrs of LOFAR HBA data in co-observing with the LOFAR Two Metre Sky Survey \citep[LoTSS;][]{Shimwell_2017,Shimwell_2019}. The new radio images were combined with archival Giant Metrewave Radio Telescope (GMRT) and {\it Chandra} observations to investigate the origin of the MH.\\

Interestingly, 2A0335+096 also hosts the head-tail radio galaxy $\text{GB6 B0335+096}$ ($z=0.038$), an FR I galaxy \citep[][]{Fanaroff-Riley_1974} with an extended radio tail \citep[][]{Patnaik_1988,Sebastian_2017}. Tailed FR I radio galaxies are characterized by  large-scales, low-brightness emission bent by the environmental pressure (thermal or ram-pressure) in the same direction, forming structures similar to tails. These radio galaxies are classified in two classes: narrow-angle tailed sources (NAT), with small angles between the tails \citep[][]{O'Dea_1985}, and wide-angle tailed sources (WAT), with a larger angle between the tails \citep[e.g.,][]{Miley_1980,Feretti_2002}. Based on ROSAT observations of head tail radio sources (including GB6 B0335+096), \citet[][]{Edge_1995} studied the connection between the enhanced X-ray emission and radio core emissions, concluding that the X-ray enhancement likely results from nuclear emission rather than substructure in the extended cluster gas. The study of radio tails can provide useful insights into the properties of the ICM, specifically on the interplay between the relativistic jets and the thermal ICM \citep[e.g.,][]{Bex_2017,Muller_2021}, as well as allowing us to probe the CRe life-cycle in the ICM. For the case of 2A0335+096, the new LOFAR data permit us to ascertain if turbulence can extend outside the cool core and interact with the CRe released by the galaxy. \\

In this work we used the standard concordance 
cosmology parameters $H_0 = 70 \, \rm km \, s^{-1} \, Mpc^{-1}$, ${\Omega}_M=0.3$ and ${\Omega}_{\Lambda}=0.7$. At the redshift of 2A0335+096, $z$=0.035, this yields\footnote{{\tt http://www.astro.ucla.edu/$\#$7Ewright/CosmoCalc.html}} 1$''$= 0.694 kpc and a luminosity distance of 145 Mpc.

\section{Data processing and analysis}
\subsection{Radio data}
\begin{figure*}[t!]
\centering
 \includegraphics[width=\linewidth]{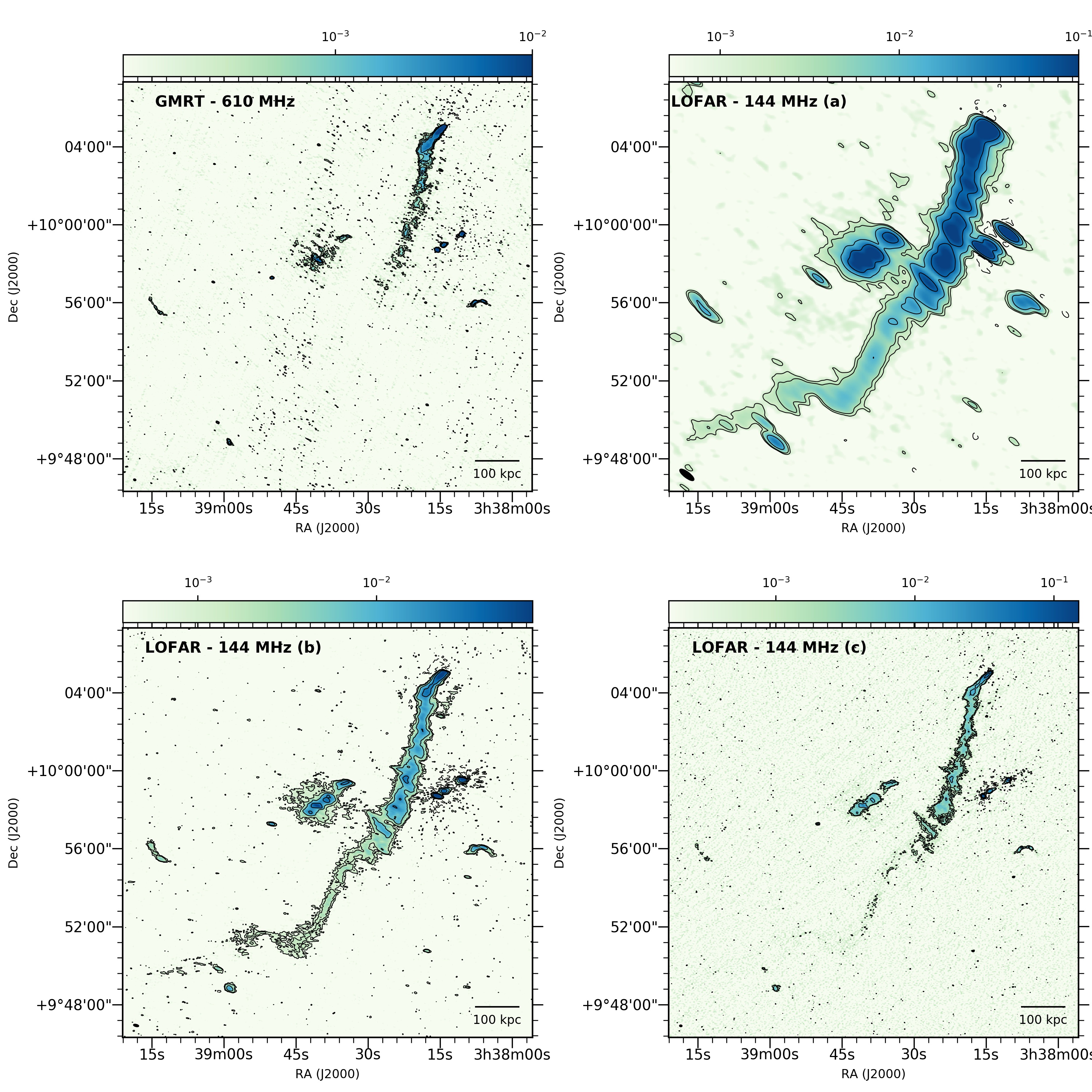}
 \caption{\label{2a0335_radio}Multi-frequency view of 2A 0335+096. Top left: GMRT image at 610 MHz, $\sigma$=84 $\mu$Jy beam$^{-1}$, resolution $6''\times4''$; Top right: LOFAR low-resolution image, $\sigma$=517 $\mu$Jy beam$^{-1}$, resolution $52''\times16''$; Bottom left: LOFAR mid-resolution image, $\sigma$=231 $\mu$Jy beam$^{-1}$, resolution $14''\times5''$; Bottom right: LOFAR high-resolution image, $\sigma$=400 $\mu$Jy beam$^{-1}$, resolution $6''\times4''$; We report the -3, 3, 6, 24, 96$\sigma$ levels in the LOFAR images and the -3, 3, 6, 12, 24$\sigma$ in the GMRT image. The maps are reported in units of Jy.}
\end{figure*}
We present the analysis of the LOFAR observation LC11$\_$014 (PI Ignesti) in the 120-168 MHz band. We reduced the dataset using the direction-dependent calibration and imaging pipeline {\scshape ddf-pipeline}\footnote{{\tt https://github.com/mhardcastle/ddf-pipeline}} v. 2.2 . This pipeline was developed by the LOFAR Surveys Key Science Project and corrects for ionospheric and beam model errors in the data. The latest version of the pipeline is described in \citet[][]{Tasse_2021}. The entire data processing procedure makes use of {\scshape prefactor}  \citep[][]{vanWeeren_2016,Williams_2016,deGasperin_2019}, {\scshape killMS} \citep[][]{Tasse_2014,Smirnov_2015} and {\scshape DDFacet} \citep[][]{Tasse_2018} and results in a thermal noise-limited image of the entire LOFAR field-of-view. To further refine the processing, we then performed additional phase and amplitude self-calibration cycles in a small region centered on the source, where the direction-dependent errors are assumed to be negligible \citep[see][for further details on this procedure]{vanWeeren_2021}. \\

We produced images at a central frequency of 144 MHz with different resolutions by using {\scshape WSClean} v2.6 \citep[][]{Offringa_2014} and using several different Briggs weightings \citep[][]{Briggs_1995}, tapering configurations, and multi-scale cleaning \citep[][]{Offringa_2017}. An inner uvcut of 80$\lambda$, corresponding to an angular scale of 43$'$, was applied to the data to remove the shortest spacings where calibration is more challenging. We corrected for the systematic offset of the LOFAR flux density scale produced by inaccuracies in the LOFAR HBA beam model by multiplying it with a factor of 0.8284 . This aligns it with the flux-scale of the point-sources mapped in the first alternative data release from the TIFR GMRT Sky Survey \citep[TGSS-ADR1,][]{Intema_2017}. Following LoTSS, we adopt a conservative calibration error of 20$\%$, which dominates the uncertainties on the LOFAR flux densities. The resulting images, with resolutions between 6 and 50 arcseconds, are shown in Figure \ref{2a0335_radio}. 2A0335+096 has a low declination of +10$^{\circ}$ that resulted in a noise above 200 $\mu$Jy beam$^{-1}$, which is consistent with that archived at higher declination \citep[e.g.,][]{Shimwell_2019} when accounting for the projection of the LOFAR stations.
 We note the presence of artifacts surrounding the bright sources, such as the AGN of GB6 B0335+096 (Figure \ref{2a0335_radio}, panel b), which are due to imperfections in our calibration.\\

For a spectral analysis we make use of a GMRT image of 2A0335+096 that we made from an archival observation at 610 MHz (cycle 17, project 17$\_$016, PI Raychaudhury). The time on target is 75 mins and the data were processed using the {\scshape SPAM} pipeline \citep[see][for details]{Intema_2009,Intema_2017}. We produced new images at 610 MHz by using {\scshape WSClean} (Figure \ref{2a0335_radio}, panel a). For the GMRT images we adopted a calibration error of 5$\%$ \citep[][]{Chandra_2004}.

\subsection{X-ray data}
\begin{figure*}[h!]
\begin{center}

\includegraphics[width=.7\textwidth]{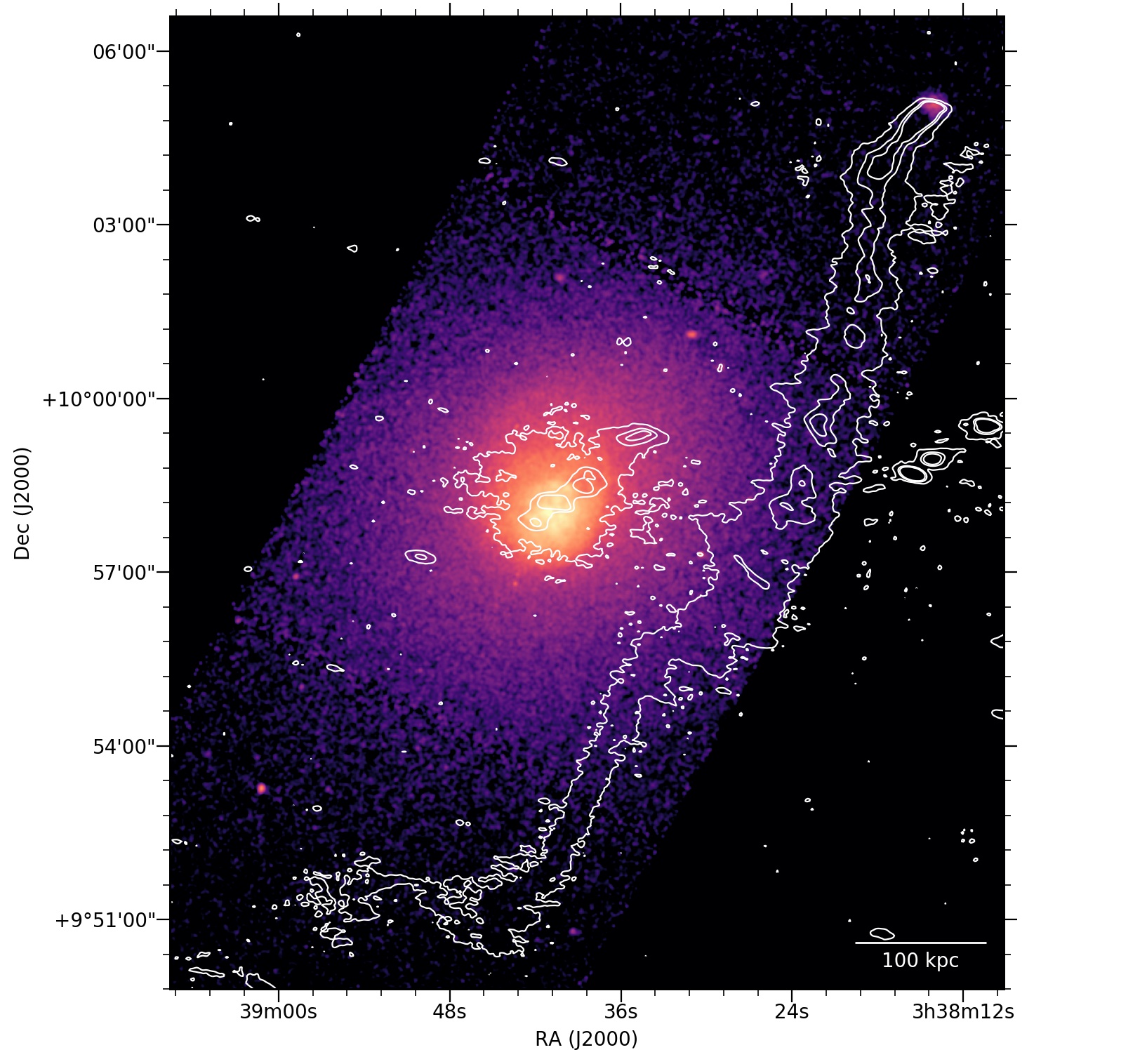}
\begin{multicols}{2}
\includegraphics[width=\linewidth]{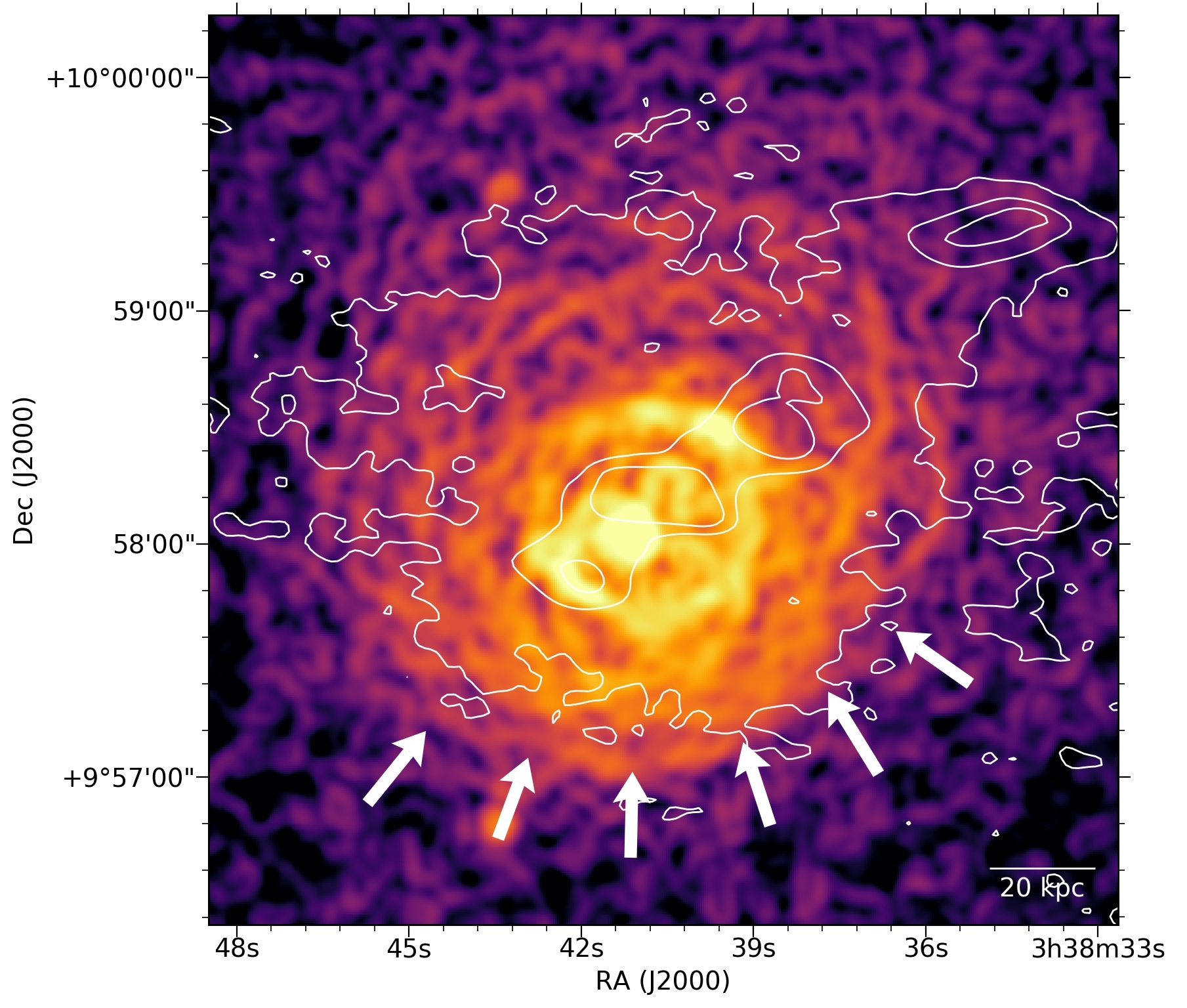}\par
\includegraphics[width=\linewidth]{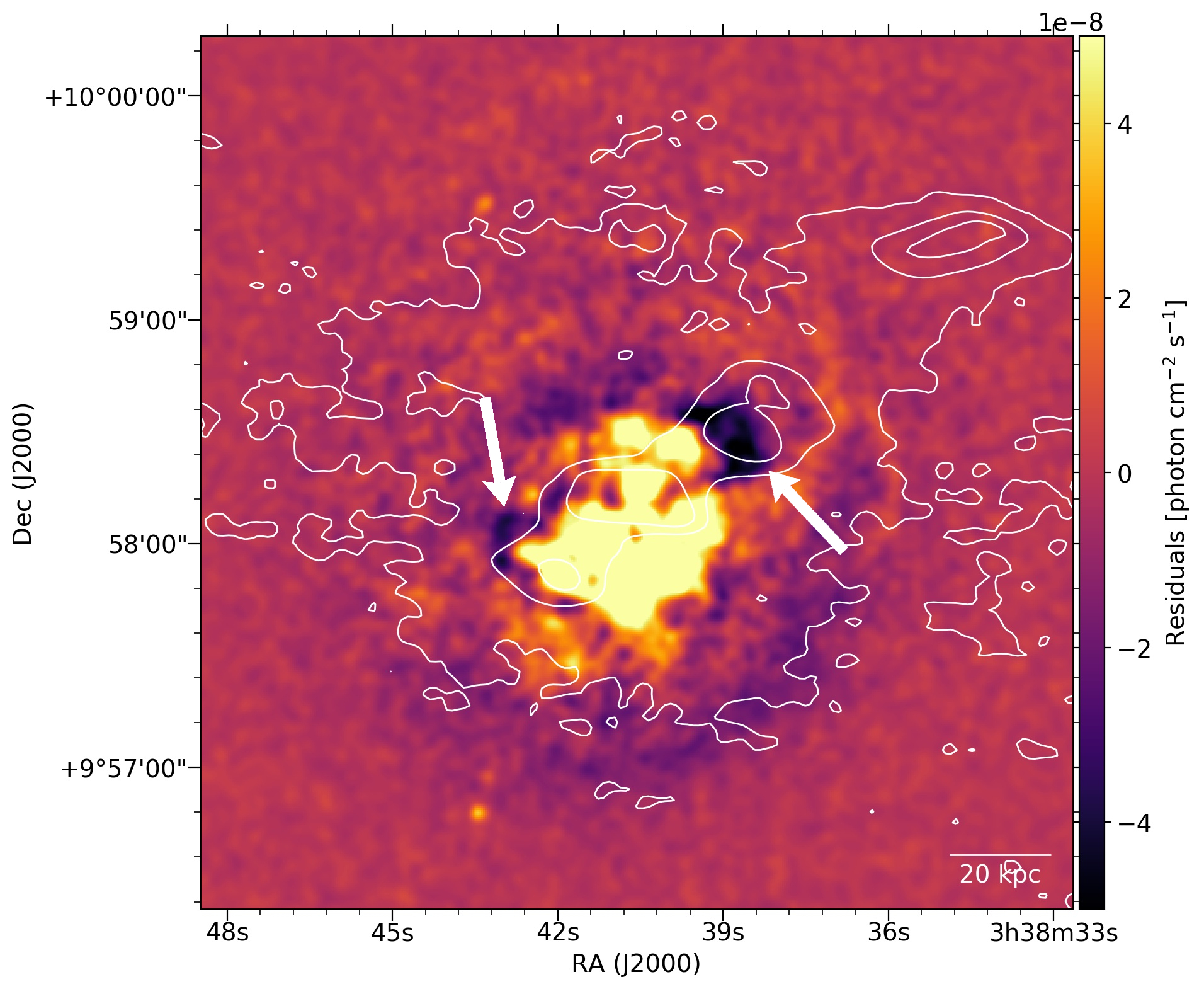}\par
\end{multicols}
\end{center}
\caption{\label{2a0335_X}  Top: LOFAR contours at 144 MHz of the mid-resolution image (panel b, Figure \ref{2a0335_radio}, $\sigma$=231 $\mu$Jy beam$^{-1}$, resolution $14''\times5''$ ) at the 3, 24, 96$\sigma$ levels on top of the background-subtracted, exposure-corrected {\it Chandra} image in the $0.5-2.0$ keV band; Bottom-left:  {\scshape ggm}-processed X-ray image produced with $\sigma=4$ and the 144 MHz emission contours on top; Bottom-right: Residual unsharp-masked image obtained by combining X-ray maps smoothed with 3 and 30 pixel-wide Gaussians with the 144 MHz emission contours on top. The white arrows in the bottom panels point to the spiral pattern and the cavities described in Section 3.1 .}
\end{figure*}
In order to study the interplay between the relativistic plasma and the thermal ICM, we produced new images in the 0.5-2.0 keV band to search for substructures in the ICM associated with the radio emission (in form of cavities or edge in the surface brightness) and to study the spatial correlation between radio and X-ray emissions (see Section \ref{ptp_sec}). For this purpose, we selected the {\it Chandra} observation 7939 (PI Sanders), which is currently the deepest archival observation (total exposure time 49.5 ks). We reprocessed the {\it Chandra} datasets with CIAO 4.10 and CALDB 4.8.1 to correct for known time-dependent gain and for charge transfer inefficiency. In order to filter out strong background flares, we also applied screening of the event files\footnote{{\tt http://cxc.harvard.edu/ciao/guides/acis\_data.html}}. For the background subtraction, we used the CALDB $``$Blank-Sky" files normalized to the count rate of the source image in the 10-12 keV band. The background-subtracted, exposure corrected image in the $0.5-2.0$ keV band is presented in the top panel of Figure \ref{2a0335_X}. We further processed the image to investigate the presence of sub-structures in the ICM. We produced a new image with the {\scshape ggm} filter \citep[][]{Sanders_2016}, which emphasizes the gradients in the surface brightness, by setting the parameter $\sigma=4$ (Figure \ref{2a0335_X}, bottom-left panel). Finally, to highlight the presence of cavities in the ICM, we produced also an unsharp-masked image by combining X-ray maps smoothed with a 3 and 30 pixel-wide Gaussians (Figure \ref{2a0335_X}, bottom-right panel).
\subsection{Spectral index map}
\label{spettro}
\begin{figure}[t!]
\centering
\includegraphics[width=\linewidth]{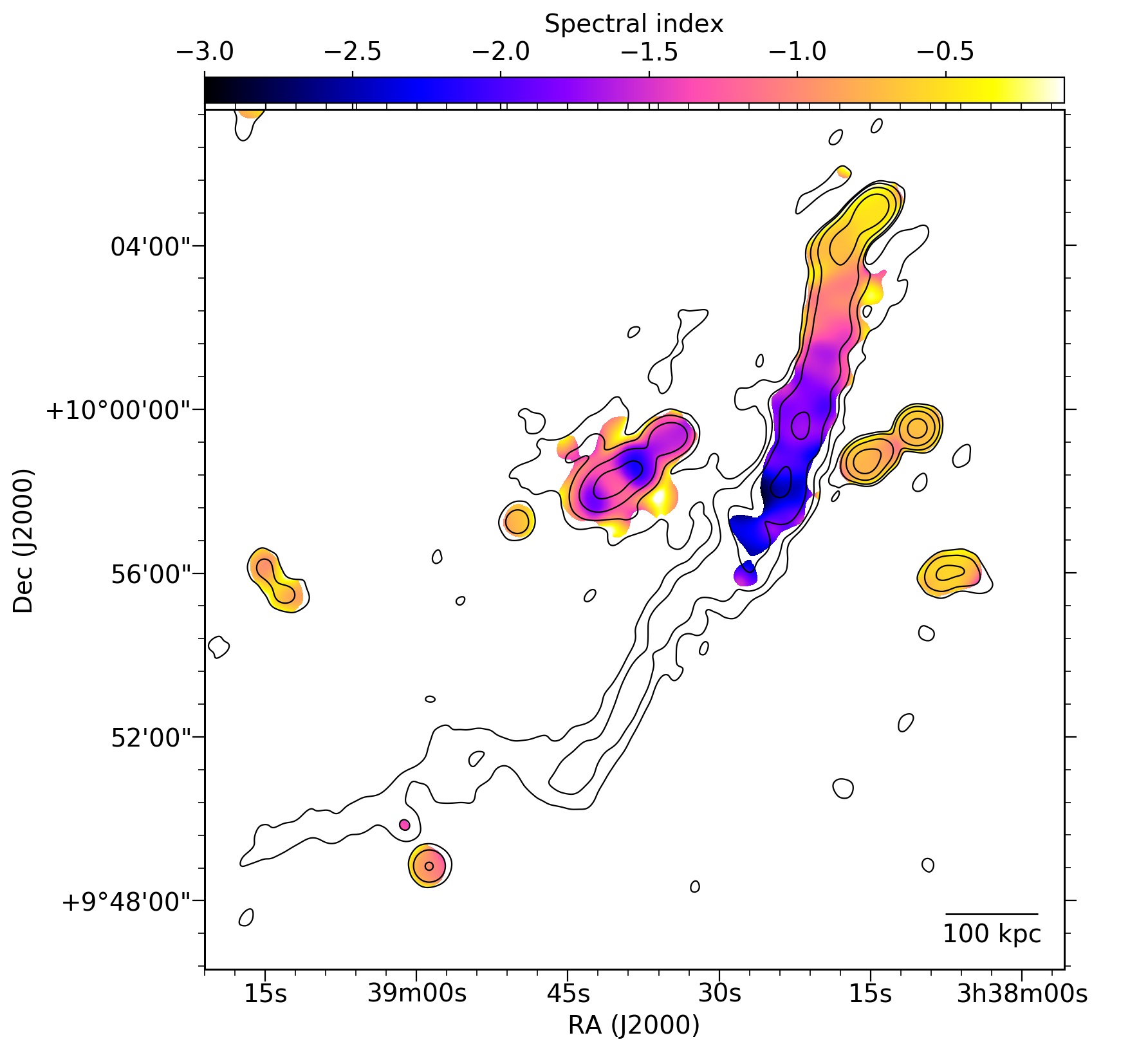}\\
\includegraphics[width=\linewidth]{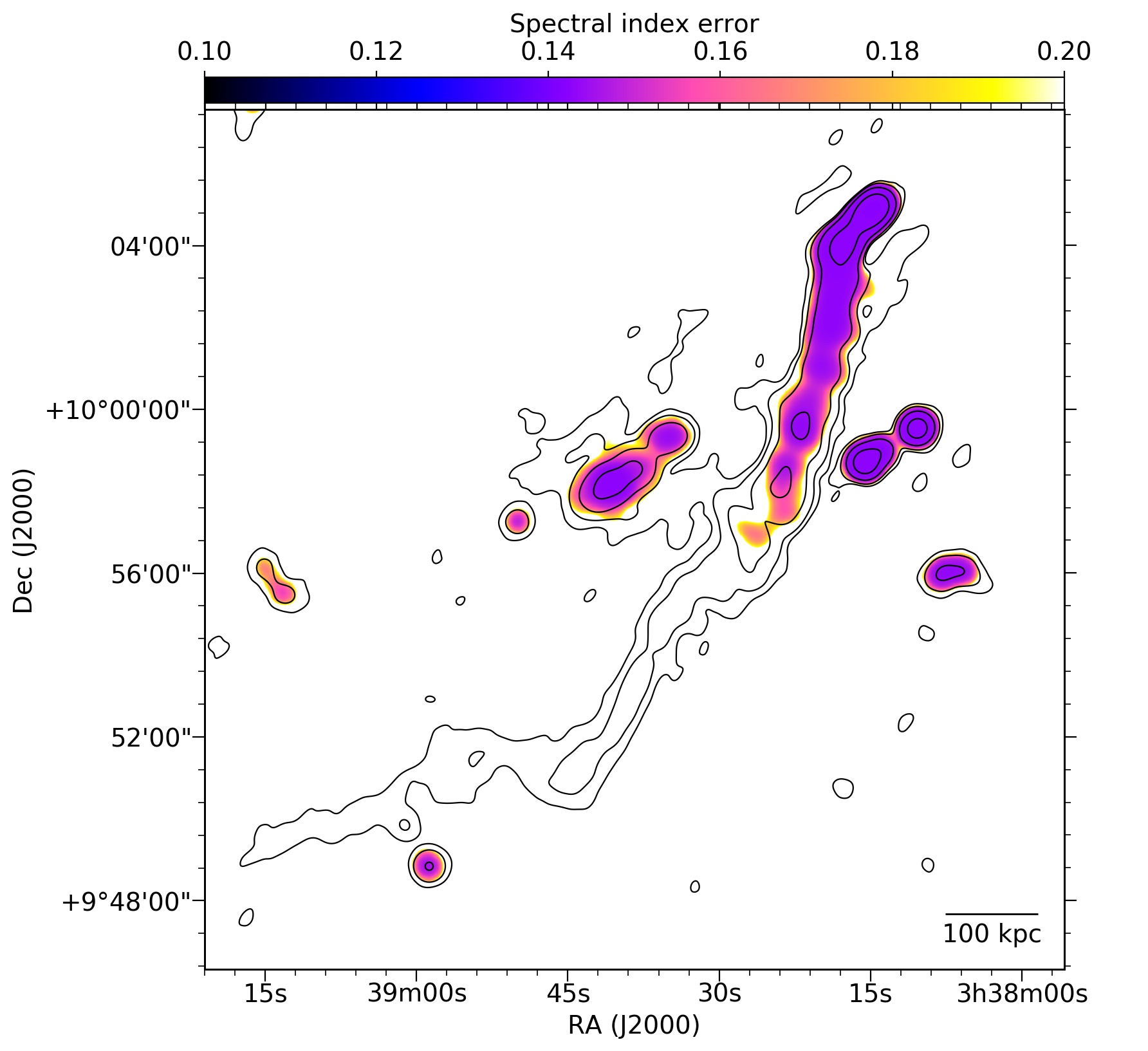}
\centering
\caption{\label{2a0335_spectral} Spectral index map between 144 and 610 MHz (top) and the corresponding error map (bottom). We report the 3, 12, 48, 192$\sigma$ contours of the 144 MHz images ($\sigma$=750 $\mu$Jy beam$^{-1}$) produced to map the spectral index.}
\end{figure}

We combine the 144 and 610 MHz images to map the spectral index of the central source and the head tail galaxy. To do this we first produced images at both frequencies with the same $uv$-range (200-54000 $\lambda$) to compare fluxes from data with the same range of spatial scales. Then we smoothed both images to a resolution of 30$''$, which was close to the resolution of the LOFAR image produced with the aforementioned $uv$-cut, and we regridded the GMRT image to match the LOFAR one to be sure that each pixel covers the same angular region. The final noise levels of the low-resolution images are 750 $\mu$Jy beam$^{-1}$ for the 144 MHz image and 410 $\mu$Jy beam$^{-1}$ for the 610 MHz one. The two images were compared on a pixel-by-pixel basis and the spectral index $\alpha$ was computed as:
\begin{equation}
 \alpha=\frac{\text{log}\left(\frac{S_{610}}{S_{144}}\right)}{\text{log}\left(\frac{610}{144}\right)}\pm\frac{1}{{\text{log}\left(\frac{610}{144}\right)}}\sqrt{\left(\frac{\sigma_{S, 610}}{S_{610}}\right)^2+\left(\frac{\sigma_{S,144}}{S_{144}}\right)^2}\text{ ,}
 \label{alpha}
\end{equation}
where $S$ and $\sigma_S=\sqrt{(f\cdot S)^2+\sigma^2}$ (where $f$ is the flux density scale uncertainty and $\sigma$ is the noise level) are the flux densities and the corresponding errors at the two frequencies, 144 and 610 MHz.\\

We computed the spectral index (and the corresponding error) for each pixel with a surface brightness above the 3$\sigma$ noise level. 
The results are shown in Figure \ref{2a0335_spectral}, where we present the spectral index map and the corresponding error map.

\subsection{Point-to-point analysis}
\label{ptp_sec}
\begin{figure}
\centering
\includegraphics[width=\linewidth]{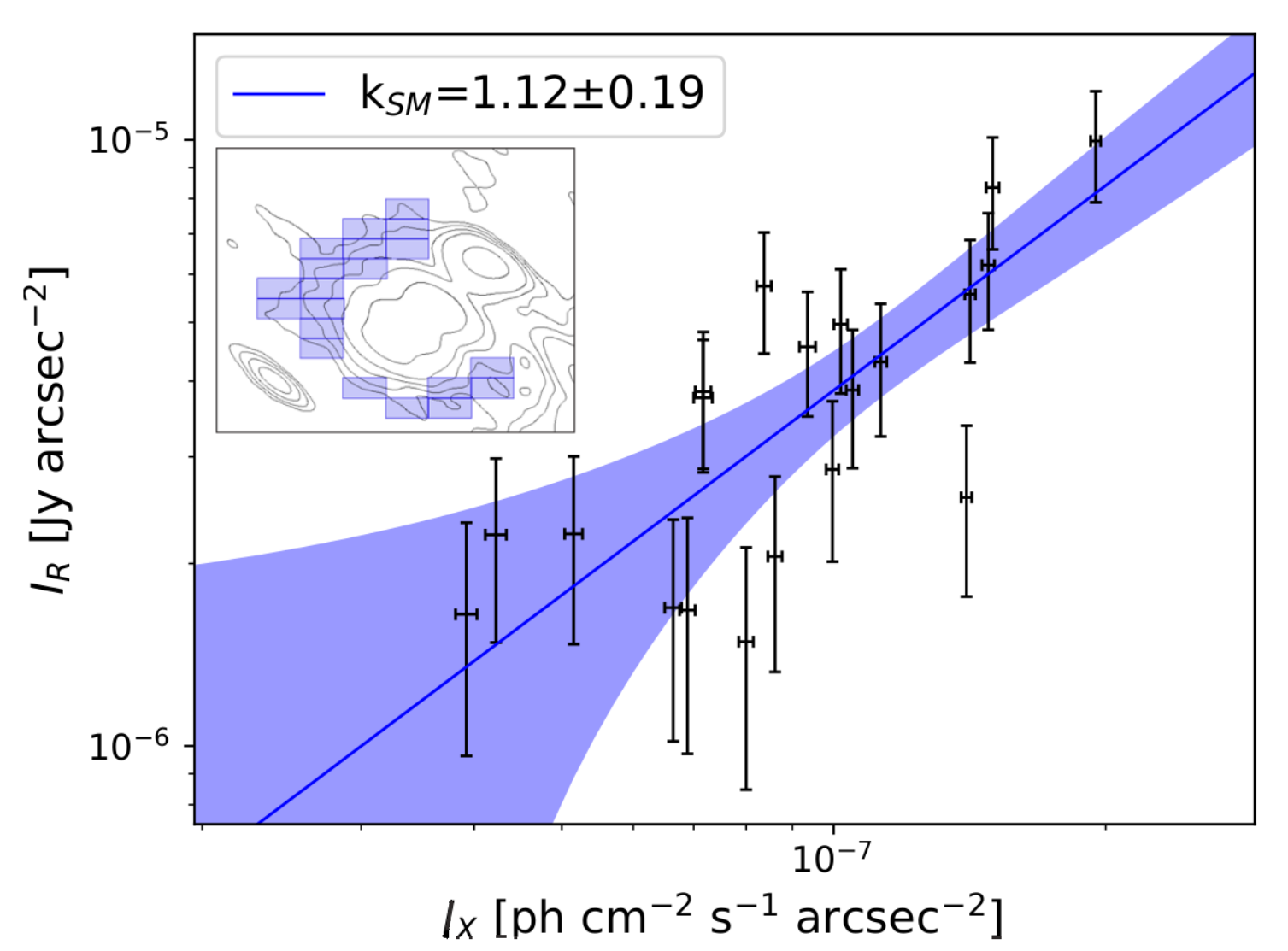}\
\includegraphics[width=\linewidth]{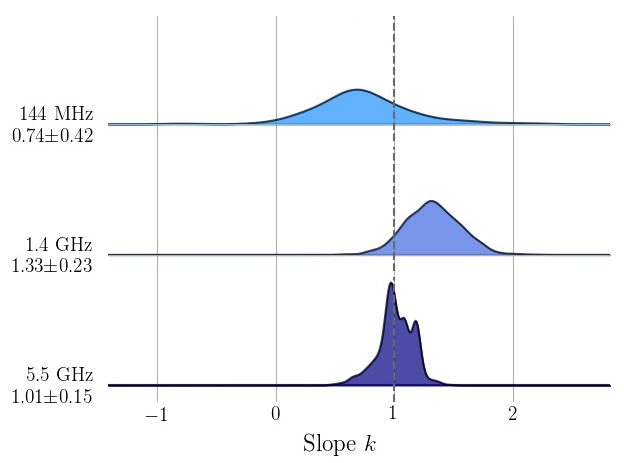}

\caption{\label{ptp} Top: result of the single-mesh point-to-point analysis. The slope of the best-fitting power-law $I_\text{R}\propto I_\text{X}^k$ is reported in the legend, and the insert presents the sampling grid overlaid on the low-resolution 144 MHz map (Figure \ref{2a0335_radio}). The blue-shaded region indicate the 90$\%$ confidence interval of the fit. Bottom: Slope index distributions produced by the MCptp analysis at 0.144 MHz, 1.4 and 5.5 GHz \citep[from][]{Ignesti_2020a}. The mean and standard deviation of each distribution is shown in the labels.}
\end{figure}
The resolution of the LOFAR data permitted us to search for a spatial correlation between the MH radio, $I_\text{R}$, and ICM X-ray, $I_\text{X}$, surface brightness. To asses this we performed a point-to-point (ptp) analysis with the Point-to-point TRend EXtractor \citep[PT-REX,][]{Ignesti_2021}. At higher frequency, the MH in 2A0335+096 exhibits a strong correlation between $I_\text{R}$ and $I_\text{X}$, in the form of $I_\text{R}\propto I_\text{X}^k$ where $k=1.33\pm0.23$ and $1.01\pm0.15$ at 1.4 and 5.5 GHz respectively. This is in agreement with the others MHs  \citep[][]{Ignesti_2020a}. This previous analysis was done at a resolution of the order of 18-20 $''$. To sample the 144 MHz radio surface brightness with a comparable resolution we used the low-resolution LOFAR image ($52''\times16''$, Figure \ref{2a0335_radio}). In order to focus on the MH, we masked the central AGN, the lobes, and the putative bridge between the MH and the radio tail.\\

At lower frequency, the correlation is less strong. A a single-mesh (SM) analysis resulted in a super-linear scaling ($k_\text{SM}=1.12\pm0.19$, Figure \ref{ptp}, top panel), whilst the Monte Carlo ptp analysis \citep[we refer to][for further details]{Ignesti_2021}, revealed that, when testing 100 different grid configurations, the correlation can assume a broad distribution of slopes (Figure \ref{ptp}, bottom panel), although the preference was for $k=0.72\pm0.42$. A sub-linear relation would be typical of giant radio halos \citep[e.g.,][]{Ignesti_2020a,Bruno_2019,Rajpurohit_2021,Biava_2021}. A similar slope has already been reported for the MH in the Phoenix cluster \citep[$k=0.84\pm0.23$, ][]{Timmerman_2021}. However, in the case of 2A0335+096 our fit is poorly-constrained, likely due to the presence of the AGN lobes and the radio tail that, even though they were masked, are bright enough to contaminate the surrounding diffuse emission resulting in a off-center increase of $I_\text{R}$ (i.e., a flattening of the spatial correlation). Hence, at present, we are unable to reliably infer further information on the CRe dynamics from the point-to-point correlation. On the other hand, it shows how the SM ptp analysis could be biased when performed on sources with complex morphology or multiple components.

\section{Results and discussion}
\begin{figure}
 \centering
 \includegraphics[width=.44\textwidth]{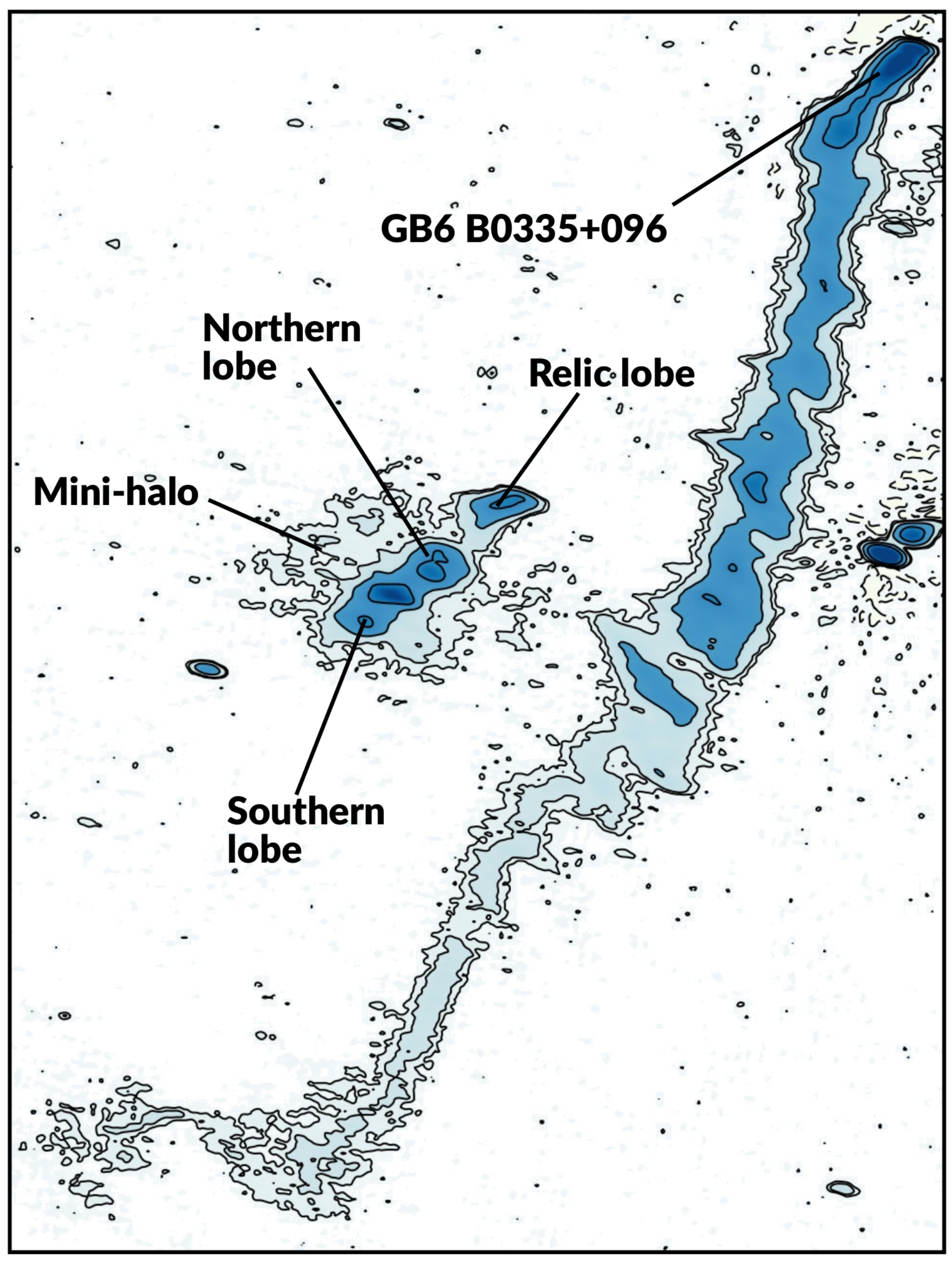}
 \caption{LOFAR contours of the mid-resolution image (panel b, Figure \ref{2a0335_radio}, $\sigma$=231 $\mu$Jy beam$^{-1}$, resolution $14.43''\times5.05''$) where we label the components of the central source as given in the text. \label{2a0335_lab}}
\end{figure}

The new LOFAR observation allows us to explore the impact of ICM turbulent re-acceleration on the lifecycle of the low-energy CRe. Here we analyze the central source, which consists of several distinct  components (Figure \ref{2a0335_lab}), and the radio galaxy GB6 B0335+096. \
\subsection{The central radio source}
LOFAR's sensitivity to large-scale emission provides us with a detailed view of the structure of the central radio source that we can explore at various resolutions. The mid-resolution LOFAR image (Figure \ref{2a0335_radio}, panel b, resolution $14''\times5''$) shows the bright central emission of the radio galaxy and the northern, southern and relic lobes embedded within the MH. The {\it Chandra} X-ray image in 0.5-2.0 keV shows the cavities produced by the radio-mode AGN feedback at the center of the cluster. In particular, the unsharp-masked image (Figure \ref{2a0335_X}, bottom-right panel) highlights that the cavity reported by \citet[][]{Mazzotta_2003} is coincident with the northern radio lobe, while the opposite radio lobe is located close to the southern cavity. In the high-resolution LOFAR image (Figure \ref{2a0335_radio}, panel c, resolution $6.13''\times3.82''$), due to lower surface brightness sensitivity, we detect only the AGN lobes, plus the relic lobe in the north west \citep[][]{Giacintucci_2019}. Finally, at the lowest resolution (Figure \ref{2a0335_radio}, panel a, resolution $52''\times16''$), where the surface brightness sensitivity is highest, we observe the previously discussed aspects but also see a putative bridge of radio emission between the central source and the radio tail. The apparent discrepancy between the spectral indices at the two ends of the putative radio bridge (Figure \ref{2a0335_spectral}) challenges the idea that the same relativistic plasma may extend from the central source to the radio tail and, hence, suggests that this feature is a blend of their emission produced by the resolution of the image. \\

Here we focus on the study of the MH. A detailed analysis of a LOFAR observation of the central AGN and its lobes can be found in \citet[][]{Birzan_2020}. In the low-resolution LOFAR image (Figure \ref{2a0335_radio}, panel a), we estimated the MH radius following \citet[][]{Cassano_2015}, i.e. $R_\text{MH}=\sqrt{R_\text{min}\times R_\text{max}}=\sqrt{75''\times127''}=97.6''\simeq67$ kpc, where $R_\text{min}$ and $R_\text{max}$ are the radii of the circular regions centered on the AGN that inscribes and circumscribes, respectively, the MH. This is consistent with the observed radius at 1.4 GHz \citep[70 kpc][]{Giacintucci_2014a}, and generally smaller than other MHs ($100-150$ kpc). However, due to the high noise of the images and the presence of a bright, extended radio galaxy at the center of the MH, we cannot be certain that we are able to measure the full extent of the system.\\

 In the mid- and low-resolution images we measured an integrated flux density of the central region (mini-halo plus central radio galaxy and lobes) at 144 MHz of 1.15$\pm$0.23 Jy within the 3$\sigma$ contours and bounded by the 6$\sigma$ contour in the direction of the radio tail. In order to estimate the flux density of the diffuse emission only, we first subtracted the contributions of the AGN lobes (0.75$\pm$0.15 Jy) and of the north-west relic lobe (0.17$\pm$0.03 Jy), both measured within the 3$\sigma$ contour of the high-resolution image (Figure \ref{2a0335_radio}, panel c). The resulting net integrated MH flux density is 0.23$\pm$0.05 Jy, where the error is computed by accounting for the $20\%$ flux density scale uncertainty as well as the error from the noise and the net area ( $\sigma_S=[(f\cdot S)^2+\sigma^2]^{1/2}$). This entails an average MH surface brightness of 1.23$\pm$0.27 mJy beam$^{-1}$ with a beam size of $14''\times5''$. Then we estimated the 'missing' MH flux density corresponding to the AGN and relic regions in the mid-resolution image  by multiplying the MH average surface brightness by their areas (estimated as the area encompassed by the 12$\sigma$ contours). We computed an additional contribution of 77.7$\pm16.9$ mJy associated with the MH emission, that results in a total flux density of 387.7$\pm$53.9 mJy.\\
 
 By comparing this measure with the flux density of the MH at 1.4 GHz reported in \citet[][]{Giacintucci_2014a}, which is 21.1$\pm$2.1 mJy, we estimated a spectral index $\alpha=-1.2\pm0.1$ between 144 MHz and 1.4 GHz. This entails a radio power at 144 MHz of $P=4\pi D_L^2 S(1+z)^{1+\alpha}=9.9\times10^{23}$ Watt Hz$^{-1}$, where $D_L=145$ Mpc is the cluster luminosity distance. We did not compute the flux density and radio power at 610 MHz due to the low quality of the current observation.\\

 We note that the MH in 2A0335+096 is a low mass, local example of this class in terms of radio power at 1.4 GHz, as well as the size \citep[][]{Giacintucci_2014a, Giacintucci_2017}. Unfortunately, the the resolved MH spectral index could not be reliably mapped (Figure \ref{2a0335_spectral}) because of the insufficient sensitivity of the GMRT observation. We were however able to measure the spectral index of the AGN lobes and the relic lobe, which show steep spectra with  $\alpha\simeq-1.5\pm0.1$. Interestingly, the {\it Chandra} {\scshape GGM} image presented in the bottom-left panel of Figure \ref{2a0335_X} shows that the southern part of the MH is encompassed by a bright, spiral-like edge in the X-ray surface brightness. This structure corresponds with the front already reported by previous studies \citep[][]{Mazzotta_2003, Ghizzardi_2006,Sanders_2006, Sanders_2009b} and it can be interpreted as evidence of sloshing of the cool core. Interestinlgy, \citet[][]{Mazzotta-Giacintucci_2008, ZuHone_2013} discussed the possibility that, by injecting turbulence in the ICM, the sloshing could be able to originate MHs. \\
 
The new LOFAR observation allows us to study the nature of the MH and its interplay with the central AGN. The spectral index map (Figure \ref{2a0335_spectral}) show that the northern and southern AGN lobes have steep spectra ($\alpha\simeq-1.5\pm0.1$). Meanwhile, the surrounding diffuse radio emission appears distinct from this morphologically but it also has a flatter\footnote{These spectra are estimated using different bands, namely 144-610 MHz for the lobes and 144-1400 MHz for the MH. The fact that the MH results to be flatter despite the larger band suggests that its spectrum is overall flatter than the lobes one.} spectrum ($\alpha=-1.2\pm0.1$). This might suggest that the CRe which are powering the MH were not simply injected from the lobes and later propagated to larger scales without any re-acceleration because, in this case, we would expect the MH spectrum to be steeper (i.e. the CRe to be older) than the lobes. Instead, the flatter spectrum could either indicate that there is no relation between the AGN and the MH, or the action of turbulent re-acceleration  affecting the CRe lifecyle. Obvious sources of such turbulence would be the AGN mechanical feeding and sloshing, the second one supported also by the close spatial correspondance between the radio emission and the sloshing region (Figure \ref{2a0335_X},  bottom-left panel). \\

Alternatively, one might assume that the CRp generated by the AGN escape the lobes at late times \citep[e.g.,][]{Ehlert_2018} and propagate around the core generating CRe via the proton-proton cascade that, in turn, will generate diffuse radio emission. The analyzed data do not allow us to check for spectral steepening in the MH, which might test this possibility \citep[e.g.,][]{Brunetti-Jones_2014}, hence this remains a viable scenario.

\subsection{GB6 B0335+096}
\begin{figure*}[t!]
\centering
\begin{multicols}{3}
\includegraphics[width=\linewidth]{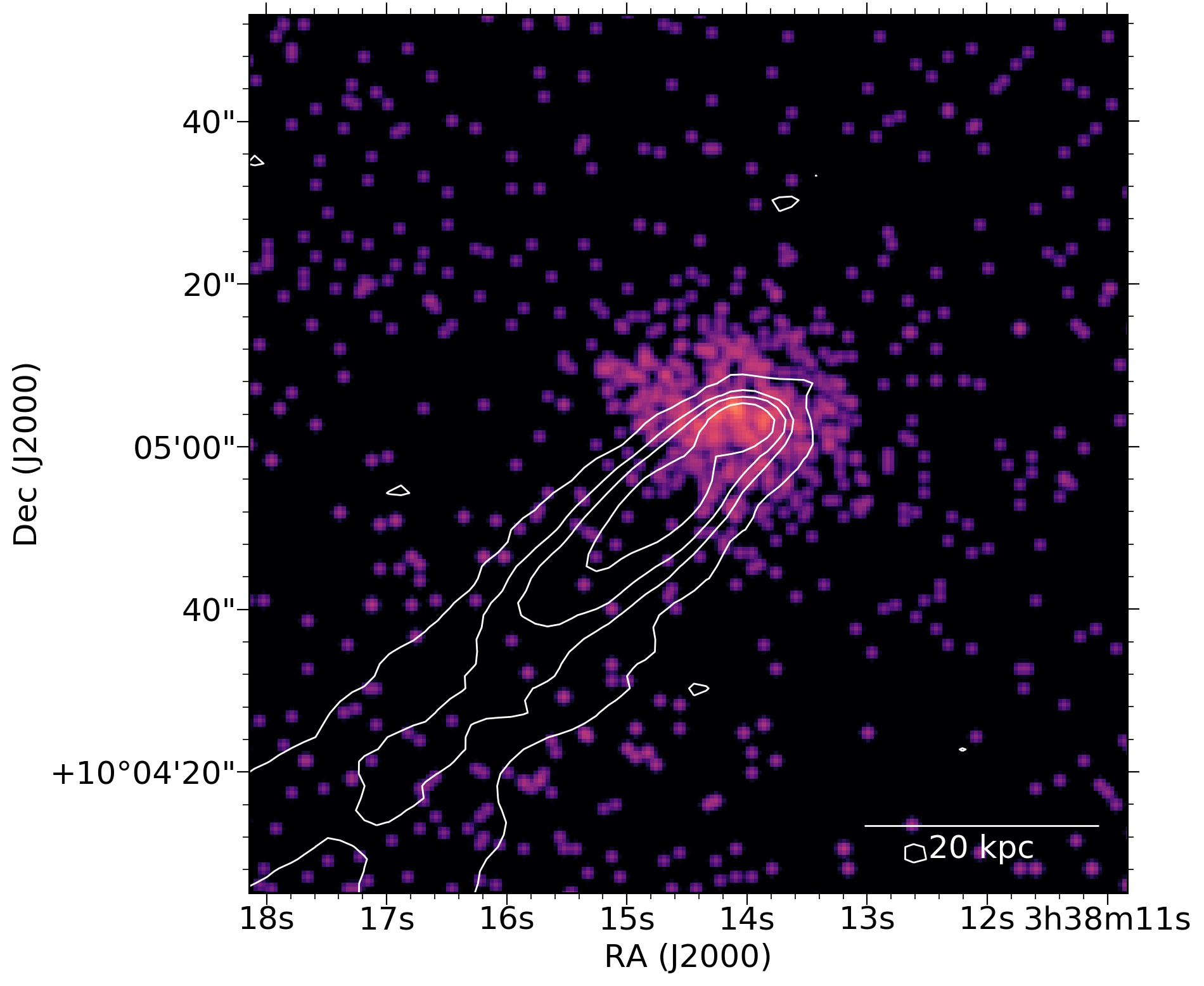}\par
\includegraphics[width=\linewidth]{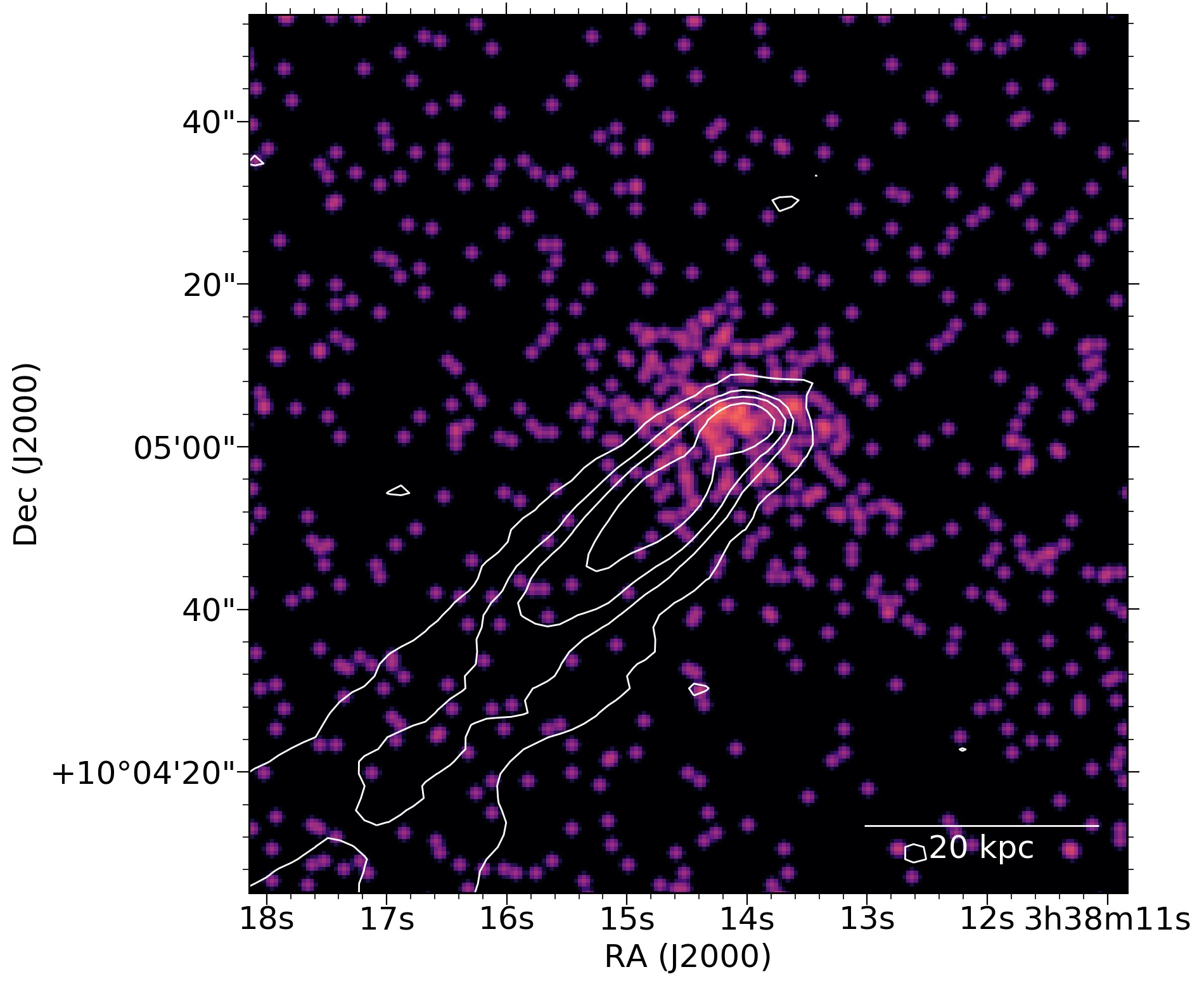}\par
\includegraphics[width=\linewidth]{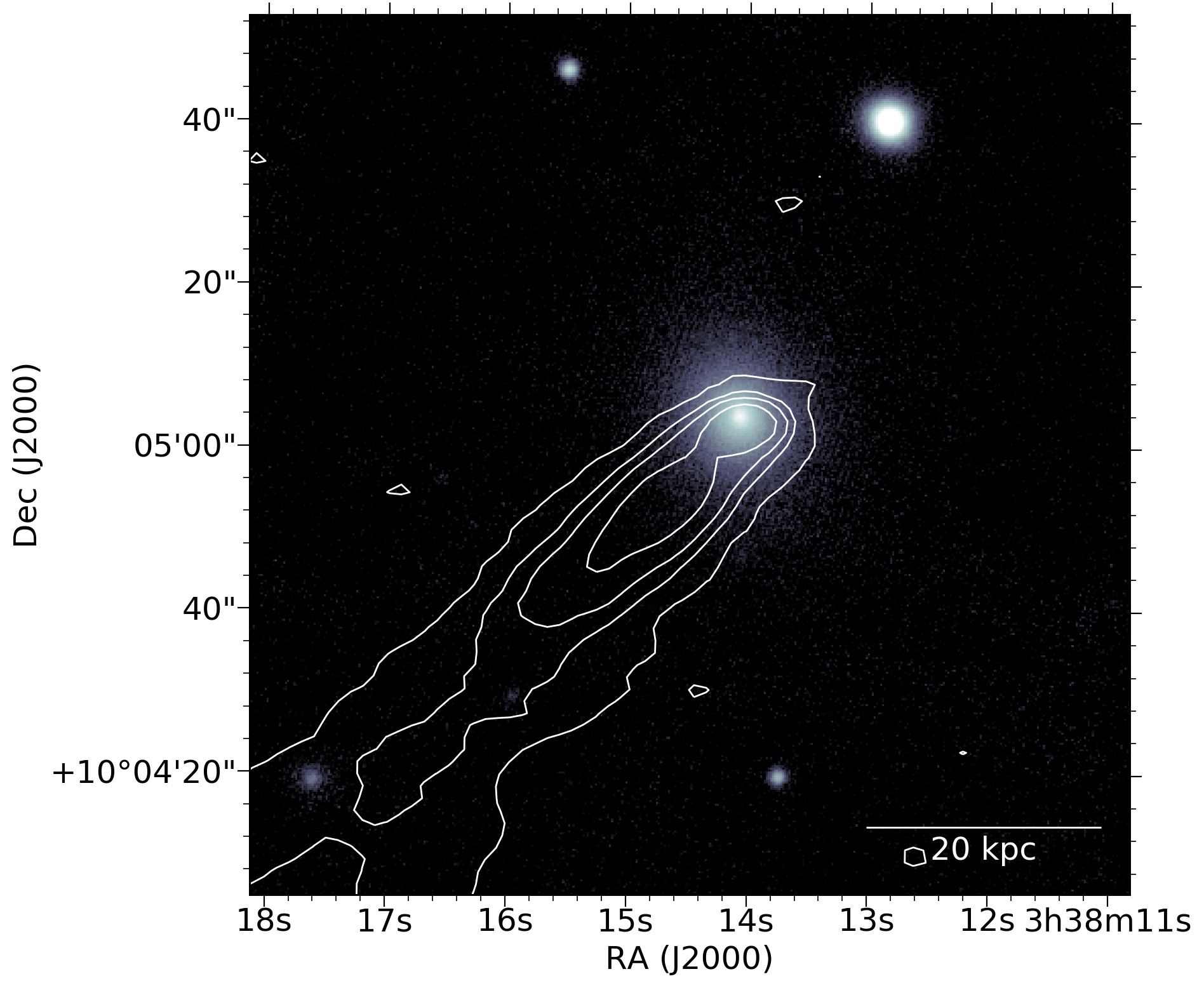}

\end{multicols}
\caption{\label{2a033_AGN}LOFAR contours of the high-resolution image at 144 MHz (panel c, Figure \ref{2a0335_radio}, $\sigma$=400 $\mu$Jy beam$^{-1}$, resolution $6.13''\times3.82''$) at the 3, 24, 48, 96$\sigma$ levels on top of the {\it Chandra} 0.5-2.0 keV (left), 2.0-7.0 (middle) and PanSTARR {\it r}-band (right) images.}
\end{figure*}

The most striking feature observed by LOFAR is the giant radio tail of GB6 B0335+096, which extends for more than 900 kpc ($\sim22'$). This is more than double than what was observed in previous works \citep[$\sim 400$ kpc][]{Patnaik_1988,Sarazin_1995,Sebastian_2017}. Interestingly, the surface brightness of the new low-frequency continuation is almost uniform up to the end of the tail. The fact that only LOFAR could observe this low-frequency continuation suggests that these structures may actually be common in many systems, thus they are going to be found more and more efficiently with the advances of the LOFAR surveys \citep[][]{Shimwell_2019,dega_2021}. The high- and mid-resolution images (Figure \ref{2a0335_radio}, panels a, b) show that the tail already bends within 100 kpc from the AGN. We detect hard X-ray 2-7 keV emission coming from the flat-spectrum core in GB6 B0335+096  (Figure \ref{2a033_AGN}), this implies that the AGN is currently active. \\

The extent of this tail is intriguing. The maximum life-time of low-energy CRe emitting at 144 MHz, which can be estimated by assuming the minimal energy loss magnetic field $B=B_\text{CMB}/\sqrt{3}=2.02$ $\mu$G (where $B_\text{CMB}=3.25(1+z)^2$ $\mu$G is the Cosmic Microwave Background equivalent field) is $\tau\simeq3\times10^{8}$ yr. Under these conditions, given an estimated velocity of GB6 B0335+096 \citep[$v_g=\sqrt{2}\times\text{line-of-sight (los) velocity}=960$ km s$^{-1}$,][]{Sebastian_2017} the expected length of a radio-emitting trail left behind would be $l\simeq v\cdot\tau\simeq300$ kpc, which is a factor $\sim3$ shorter than what we observe. Moreover, the real difference is likely larger because we can measure only the projected length of the tail, which is the lower limit of this actual length. Furthermore, the actual magnetic field differs from the minimal energy loss condition, which entails that the CR lifetime $\tau$, and hence the radio tail, would be shorter. The long radio tail of GB6 B0335+096 could therefore be explained by either a galactic velocity significantly higher than inferred from the los velocity, or an extended CRe life-time due to the action of some physical process taking place in the tail.\\ 

\begin{figure*}[t!]
\centering
\includegraphics[width=\linewidth]{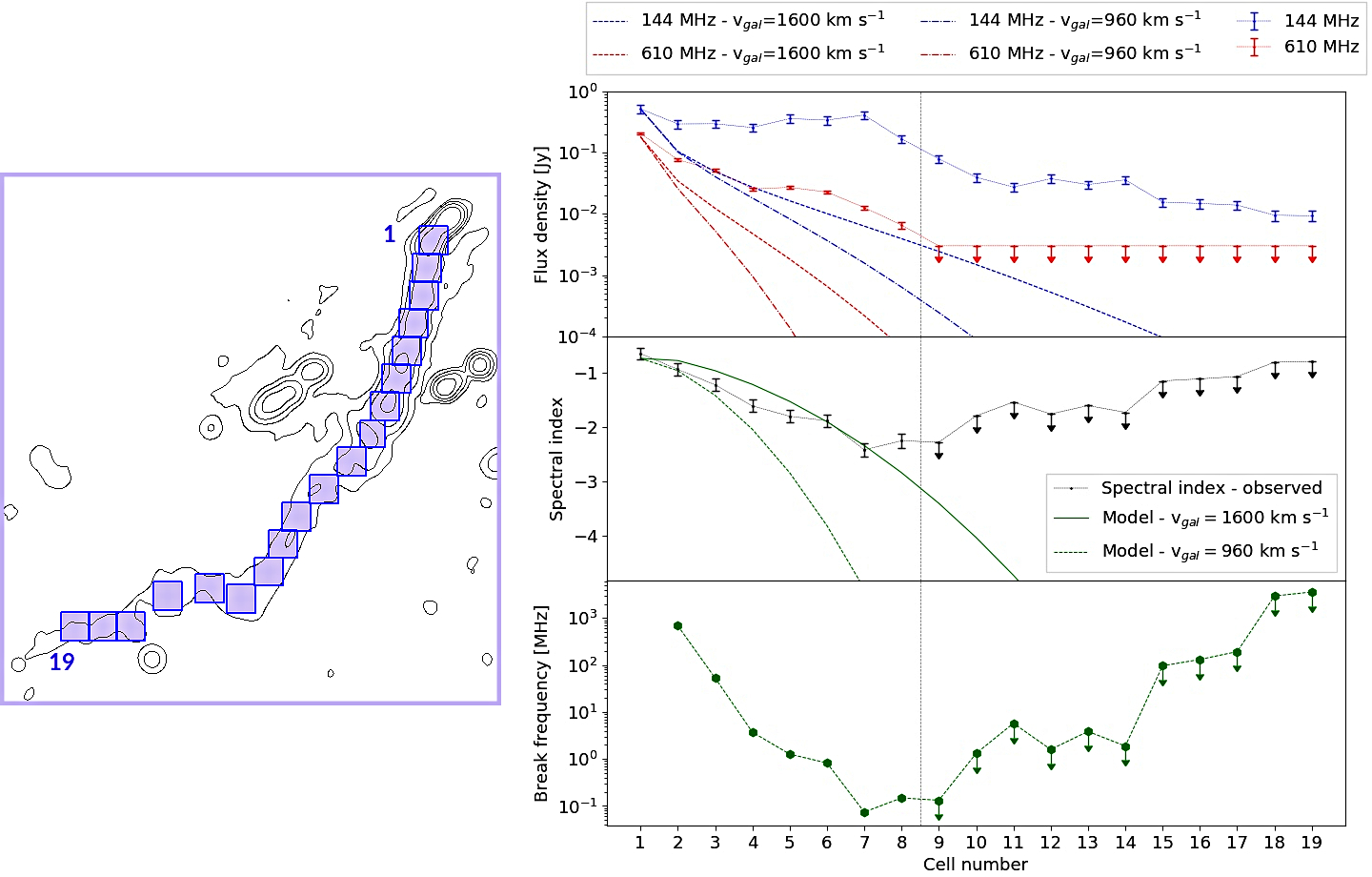}
\caption{\label{2a0335_profile} Left: 3, 48, 96, 192$\sigma$ levels of the 144 MHz radio emission ($\sigma$=750 $\mu$Jy beam$^{-1}$, resolution 30$\times$30 arcsec) with the sampling cells on top. Right: Observed and model profiles of flux density (top), spectral index (middle) and break-frequency (bottom). The upper-limits are computed based on the 2$\sigma$ level of the 610 MHz map. In the top and middle panels are reported the predictions based on the two different galactic velocities.}
\end{figure*}

 We investigated the radio tail with simple numerical calculations of the expected emission of an aging radio plasma. Our simple model allowed us to estimate the flux density profile and the corresponding decline of the spectral index along the tail. In order to compare the flux densities at 144 and 610 MHz, we used the images produced for the spectral analysis (Section \ref{spettro}) with matching uv-ranges (200-54000 $\lambda$) and resolutions (30$\times$30 arcsec). Based on the 3$\sigma$ noise level of the 144 MHz emission, we used the PT-REX code to sample the tail with 19 boxes, $60''\times60''$ in size. Then we measured the flux densities at 144 and 610 MHz in each of them, producing a profile of the emission along the tail (Figure \ref{2a0335_profile}). In this profile the 610 MHz emission is present up to box number 8. Where both LOFAR and GMRT detected the emission above the 3$\sigma$ level we computed a simple spectral index (Equation \ref{alpha}). Where we do not detect emission  at 610 MHz, we used the 2$\sigma$ level of the GMRT image to derive an upper limit for the spectral index. By assuming that the tail is parallel to the plane of the sky, our $60''\times60''$ sampling can be roughly converted to a spatial scale, which sets a lower limit for its actual length. Under these assumptions, each box is separated by 45 kpc. Assuming that the plasma left behind by the galaxy does not move with respect to the ICM, the time elapsed since the injection can be estimated as $\tau\simeq d/v$, where $d$ is the distance from the AGN and $v$ is the velocity of the plasma with respect to the AGN. For simplicity, we neglected the bulk motion of the jet and we assumed $v=v_\text{gal}$, where $v_\text{gal}$ is the galaxy velocity. It follows that the space grid defined by our sampling can be converted in a 'time grid' with the dynamic age of each chunk of the radio tail. \\

We then compare our observations with the expected synchrotron emissivity by taking into account only the radiative cooling of CRe which is dominated by synchrotron and Inverse Compton energy losses \citep[e.g.,][]{Pacholczyk_1970}. Our estimate is conservative because we do not include the energy losses due to adiabatic expansion of the plasma, which would cause the tail emissivity to steepen faster, and we assume a uniform magnetic field. We computed numerically the synchrotron emissivity spectrum $j(\xi)$, where $\xi=\nu/\nu_\text{br}$ and $\nu_\text{br}$ is the break frequency, for $\xi=0.1-10$ to properly model the spectral steepening. We assumed an injection index $\alpha=-0.6$, based on the spectral index observed at the beginning of the tail, and the minimal energy loss magnetic field $B=B_\text{CMB}/\sqrt{3}\simeq2$ $\mu$G.\\

To compare this model emissivity spectrum with our observation, we first derive the corresponding values of $\xi$ at 144 and 610 MHz by evaluating the $\nu_\text{br}$ for each box. This can be done from the radiative time of a relativistic particle population, $t_\text{rad}$, which can be expressed as:
\begin{equation}
t_r\simeq3.2\times 10^{10}\frac{B^{1/2}}{B^2+B_{\text{CMB}}^2} \frac{1}{\sqrt{\nu (1+z)}}\text{ yr,}
\label{cool_RW}
\end{equation}
where the magnetic fields are expressed in $\mu$G and the observed frequency $\nu$ is in units of MHz \citep[e.g.,][]{Miley_1980}. Under the assumption that the radiative age of the plasma coincides with the time elapsed since the injection, i.e. $t_\text{rad}\simeq \tau$, we can derive a putative $\nu_\text{br}$ for each chunk, and thus the corresponding values of $\xi$ at the two frequencies, $\xi_{144}$ and $\xi_{610}$.\\

By combining the predicted synchrotron spectrum with the time-scale described by the break frequency, we can compare the observed spectrum with the prediction in the case of pure radiative ageing. Under the assumption that each cell defines a cylindrical section of the tail with a radius of 22.5 kpc and height of 45 kpc (i.e., corresponding with the size of the sampling boxes)\footnote{Such a simple assumption does not well describe the geometry at the end of the tail where it curves toward east (cells number 10-15). However, more advanced geometrical models including the rotation of the cells would not affect affect our conclusions.} and by using the flux density at 144 MHz measured in the first cell to normalize the model spectrum $j(\xi)$, we can produce the expected flux density and spectral index profiles along the tail. \\

We compared the observed spectral index profile with extreme two predictions produced by assuming $v_\text{gal}=$ 960 and 1600 km s$^{-1}$. The first one is computed by assuming that the los velocity is $1/\sqrt{2}$ of the total velocity \citep[][]{Sebastian_2017}, whereas the second one is an upper limit estimated by forcing the model to reproduce the spectral index decline within the first half of the tail. We note that both velocities are higher than the cluster velocity dispersion $\sigma\simeq600$ km s$^{-1}$ \citep[calculated on the basis of the M$_{500}=2.27\times10^{14}$ M$_{\odot}$ and R$_{500}=0.92$ Mpc reported in][]{Giacintucci_2017}, which is not unusual for head-tail galaxies \citep[e.g.,][]{Venkatesan_1994}. Under our assumptions, the higher the velocity of the galaxy, the younger the age of the plasma in the tail, hence the slower the emissivity and the spectral index decline. Finally, to estimate the state of the plasma aging, i.e. the change of $\nu_\text{br}$, along the tail, we used the model synchrotron spectrum to sample the expected spectral index for values of $\nu_\text{br}$ within 1-5000 MHz. Then, by comparing via interpolation the observed spectral index with our predictions, we evaluated the break frequency profile along the tail. We report the observed and predicted profiles in Figure \ref{2a0335_profile}\\

Neither of the two models can reproduce the observations. They both predict that the flux density and the spectral index trends would be steeper than observed, with the emission that would fall below the detection limit within $\sim$500 kpc. In the first 500 kpc (i.e. where we still observe the 610 MHz emission), the tail shows several bright knots of emission, which can be observed at both frequencies and they appear as a change of slope of the otherwise declining trend in flux density. Because we normalize the spectrum at the flux density observed at the beginning of the tail, we expect these knots, which are not included in our simple model, to be the main cause of the difference between the observations and the predictions in the cells number $4-8$. The question is whether these features were produced by phases of more powerful AGN activity or by external processes affecting the tail. However, beyond 500 kpc (cell number 8) the observed flux density differs from the model by more than 2 orders of magnitude, with the 144 MHz emission declining dramatically more slowly than expected. The break frequency profile (Figure \ref{2a0335_profile}, bottom panel) shows a decline in the first half of the tail. Then a sudden flattening occurs in the cells 7, 8 and 9 at values of 250 MHz, followed by an apparent, inverted trend up to the end of the tail. Because the break frequency of the spectrum is an indicator of the plasma age, observing its values to be constant for $\sim 100$ kpc (and likely way further, as suggested by the flatter decline of the 144 MHz emission beyond cell number 9) indicates that in that region the radiative energy losses are balanced by some external process that is providing energy to the radiative electrons.\\

To explain the long tail, mechanisms of re-energization can be invoked. One possibility is that turbulence is generated in the tail by instabilities driven by the interplay of the tail with the external medium \citep[i.e. gentle re-acceleration, see][for similar studies]{deGasperin_2017, Wilber_2019,Botteon_2021,Muller_2021}. In this scenario, the re-energization would allow CRe to emit for longer than predicted in our simple model, thus explaining the discrepancy. Alternatively, the observed profiles could be explained by compression due to the interaction of the tail with a weak shock. This would locally flatten in the spectral index, which may be in agreement with the uniform spectral index observed in cells number $7-9$. This latter scenario would also be characterized by the presence of a discontinuity in the X-ray surface brightness in the proximity of the tail. However, searching by extracting surface brightness profiles across the tail is complicated as the tail is placed along the edge of the CCD of the {\it Chandra} observation (Figure \ref{2a0335_X},  top panel). On the other hand, a shock-like acceleration would extend the radiative life of both the high- and low- energy CRe, i.e. would result in a strong flattening of the spectral index, thus resulting in extended ($>$500 kpc) emission at both frequencies. However, the 144 MHz emission is far more extended than the 610 MHz emission. This indicates that the efficiency of the re-energizing process is a factor $\sim2$ too low (Equation \ref{cool_RW}) to compensate the energy losses observed at 610 MHz, but, still, its time-scale is consistent with the time-scale of the energy losses of the low-energy CRe emitting at 144 MHz, i.e. $\sim10^8$ yrs. Therefore we conclude that we are dealing with a low-efficient phenomenon and the ICM gentle re-acceleration appears as most likely process to power the low-energy CRe population, counter the radiative cooling and explain the presence of 144 MHz emission at such a large distance from the AGN. 

\section{Conclusions}
Due to its capability to probe the low-energy CRe, LOFAR unveiled a series of remarkable features in 2A0335+096. We focused our analysis on two aspects: the nature of the central extended radio emission and the properties of the radio tail of GB6 B0335+096. For the first aspect, we could estimate the spectral index of the diffuse radio emission, $\alpha=-1.2\pm0.1$, which is flatter than the spectral index we observe in the lobes of the central radio galaxy ($\alpha\simeq-1.5\pm0.1$). We concluded that the diffuse radio emission may be connected to the AGN (as primary source of CR), but the spectral index suggests that additional processes, either due to turbulent re-acceleration or a hadronic cascade, played a role in the origin of the MH. On the basis of the new 144 MHz images and the sloshing features highlighted by the GGM analysis, we suggest that turbulent re-acceleration may have played a role. However, with the analyzed data it is difficult to discriminate the two models. \\

For GB6 B0335+096, the new observations reveal that the radio tail is significantly extended at low frequencies ($\sim$900 kpc). This raises the questions of how the relativistic particles could still be emitting at the level observed this far from the AGN. In order to address this question, we studied the profiles of flux density and spectral index along the tail trying to reproduce them with a simple ageing model. This led to the conclusion that a re-acceleration mechanism has to be invoked to explain the observations. The fact that this continuation is revealed only at low frequency arises the possibility that such phenomena may be actually common in many systems, and hence that they will be found more and more frequently with the progress of the LOFAR surveys. Future multi-frequency and polarimetric studies will be necessary to obtain complementary constraints on the CRe properties and the magnetic field geometry and further investigate the structures of these radio tails and their interplay with the ICM.


\section*{Acknowledgments}
We thank the anonymous Referee for helping improving the presentation of this work and H. T. Intema for providing the reduced GMRT observation of 2A0335+096. AI acknowledges the Italian PRIN-Miur 2017 (PI A. Cimatti). We acknowledge financial contribution from the agreement ASI-INAF n.2017-14-H.0 (PI Moretti).
MB acknowledges support from the Deutsche Forschungsgemeinschaft under Germany's Excellence Strategy - EXC 2121 “Quantum Universe” - 390833306. RJvW and AB acknowledges support from the VIDI research programme with project number 639.042.729, which is financed by the Netherlands Organisation for Scientific Research (NWO). ACE acknowledges support from STFC grant ST/P00541/1. FdG acknowledges support from the Deutsche Forschungsgemeinschaft under Germany's Excellence Strategy - EXC 2121 “Quantum Universe” - 390833306.  GDG acknowledges support from the Alexander von Humboldt Foundation. CJR acknowledges financial support from the ERC Starting Grant `DRANOEL', number 714245. 
LOFAR \citep[][]{vanHaarlem_2013}  is the Low Frequency Array designed and constructed by
ASTRON. It has observing, data processing, and data storage facilities in several countries,
which are owned by various parties (each with their own funding sources), and that are
collectively operated by the ILT foundation under a joint scientific policy. The ILT resources
have benefited from the following recent major funding sources: CNRS-INSU, Observatoire de
Paris and Université d'Orléans, France; BMBF, MIWF-NRW, MPG, Germany; Science
Foundation Ireland (SFI), Department of Business, Enterprise and Innovation (DBEI), Ireland;
NWO, The Netherlands; The Science and Technology Facilities Council, UK; Ministry of
Science and Higher Education, Poland; The Istituto Nazionale di Astrofisica (INAF), Italy.
This research made use of the Dutch national e-infrastructure with support of the SURF
Cooperative (e-infra 180169) and the LOFAR e-infra group. The Jülich LOFAR Long Term
Archive and the German LOFAR network are both coordinated and operated by the Jülich
Supercomputing Centre (JSC), and computing resources on the supercomputer JUWELS at JSC
were provided by the Gauss Centre for Supercomputing e.V. (grant CHTB00) through the John
von Neumann Institute for Computing (NIC).
This research made use of the University of Hertfordshire high-performance computing facility
and the LOFAR-UK computing facility located at the University of Hertfordshire and supported
by STFC [ST/P000096/1], and of the Italian LOFAR IT computing infrastructure supported and
operated by INAF, and by the Physics Department of Turin university (under an agreement with
Consorzio Interuniversitario per la Fisica Spaziale) at the C3S Supercomputing Centre, Italy. We thank the staff of the GMRT that made these observations possible. GMRT is run by the National Centre for Radio Astrophysics of the Tata Institute of Fundamental Research. This research made use of Astropy, a community-developed core Python package for Astronomy \citep[][]{astropy_2013, astropy_2018}, and APLpy, an open-source plotting package for Python \citep[][]{Robitaille_2012}.

\bibliographystyle{aa}
\bibliography{bibliography.bib}
\end{document}